\documentclass{aa} 
\usepackage{multirow}
\usepackage[normalem]{ulem}

\usepackage{lineno}
\nolinenumbers
\let\linenumbers\nolinenumbers
\usepackage{graphicx}
\usepackage{txfonts}
\usepackage{capt-of}

\usepackage[colorlinks=true,linkcolor=blue,citecolor=blue,urlcolor=blue]{hyperref}

\begin{document}

\title{Prominence signatures in the Fraunhofer G-band}
\subtitle{Testing ionization memory with multi-line prominence diagnostics}

 \author{A.G.M. Pietrow\inst{1} \and  H. Balthasar\inst{1} \and P. Váradi Nagy\inst{2} \and R. Kamlah\inst{1} \and A. Stork\inst{1,3} \and C. Denker\inst{1} \and M. Verma\inst{1} 
     }

 \institute{\inst{1}Leibniz-Institut für Astrophysik Potsdam (AIP), An der Sternwarte 16, 14482 Potsdam, Germany\\
 \inst{2}Independent Researcher, Cluj-Napoca, Romania\\
 \inst{3}Universität Potsdam, Institut für Physik und Astronomie, Karl-Liebknecht-Straße 24/25, 14476 Potsdam, Germany\\
       \email{apietrow@aip.de}\\
      }

\date{Draft: compiled on \today}

\abstract{The Fraunhofer G-band around 4304~\AA\ is widely used as a photospheric diagnostic and is generally not expected to show signatures of chromospheric or coronal structures. However, recent amateur observations have suggested the presence of off-limb prominence emission in this spectral region.}
{We investigate the origin of the prominence emission in the G-band to determine if this is caused by methylene (CH) or other lines in this band. We also aim to test these lines for the presence of ionization memory effects in neutral lines.}
{We present a case study of two prominences, one obtained with a Solar Explorer (Sol’Ex) spectroheliograph and another with the high-resolution Fast Multi-Line Universal Spectrograph (FaMuLUS) camera system at the echelle spectrograph of the German Vacuum Tower Telescope (VTT). Line widths are measured for simultaneously observed neutral and ionized metal lines, allowing a comparison of thermal and non-thermal broadening components to see if these lines exhibit any ionization memory effects.}
{We report clear prominence emission in several metal lines within the G-band, primarily from \ion{Ti}{II} and \ion{Ca}{I} lines, while contributions from CH molecular lines are not observed. A comparison of the simultaneously observed ionized and neutral lines reveals no clear evidence for an ionization memory effect.}
{Since the prominence emission does not originate from CH lines, we will not call them ``G-band prominences'' but rather prominences in the G-band, as they are independent of the primary diagnostic in this spectral window. In addition, the absence of a clear ionization memory effect suggests that such effects may be less pronounced for weak neutral lines.}


 \maketitle

\section{Introduction}

The solar G-band is a roughly 10~\AA\ wide spectral band centered around 4304~\AA. It is named after the label given by \citet{Fraunhofer1817} in his original spectral classification. This band should not be confused with the much wider \citet{Johnson1953} photometric G band, which covers the green part of the spectrum. For this reason, this range is often specifically called the solar G-band, Fraunhofer G-band, or CH-band for the many methylidyne lines being present in this spectral range \citep[e.g.,][]{Langhans2002}.

The G-band is dominated by numerous closely spaced rotational-vibrational transitions of the CH molecule, making it a popular diagnostic in high-resolution solar imaging. In cooler regions of the photosphere, such as intergranular lanes, sunspots, and pores, the higher CH abundance leads to enhanced molecular absorption, raising the band's opacity and lowering the integrated intensity. Conversely, in hotter regions such as granule centers and magnetic flux concentrations (or bright points), the CH molecule is thermally dissociated. This reduces the band's opacity and allows for deeper and hotter layers to contribute to the emergent intensity. This mechanism enhances contrast relative to nearby continuum wavelengths \citep[e.g.,][]{Shelyag2004, Bodnarova2014, Kamlah2025}. Photospheric imprints of solar flares can also be observed in this band, albeit only very bright ``white-light flares'', which seem to likewise appear brighter than the surrounding continuum due to the dissociation of the CH molecules \citep{Isobe2007}. In stellar studies, the G-band is used as a probe for the carbon abundance \citep[e.g.,][]{Lee2008}.

In summary, the emission from the G-band generally forms in the low to mid-photosphere, with its formation height depending on the temperature of the local atmosphere. As such, it is not expected to exhibit signatures of chromospheric phenomena. Nevertheless, informal discussions of so-called ``G-band prominence'' have surfaced in amateur forums and social media groups, with the first confirmed detection being posted in August 2024 \citep{VaradiGallery2025}. 

To further investigate these features, a coordinated ``Professional--Amateur'' campaign was initiated using a Solar Explorer \citep[Sol’Ex;][]{Buil2023} spectroheliopgraph, which led to several additional observations \citep{Varadi2025}. In addition, attempts were made to capture these weak signals in narrow-band photometry, but without success.

\begin{figure*}[t]
\centering
\includegraphics[width=1\textwidth]{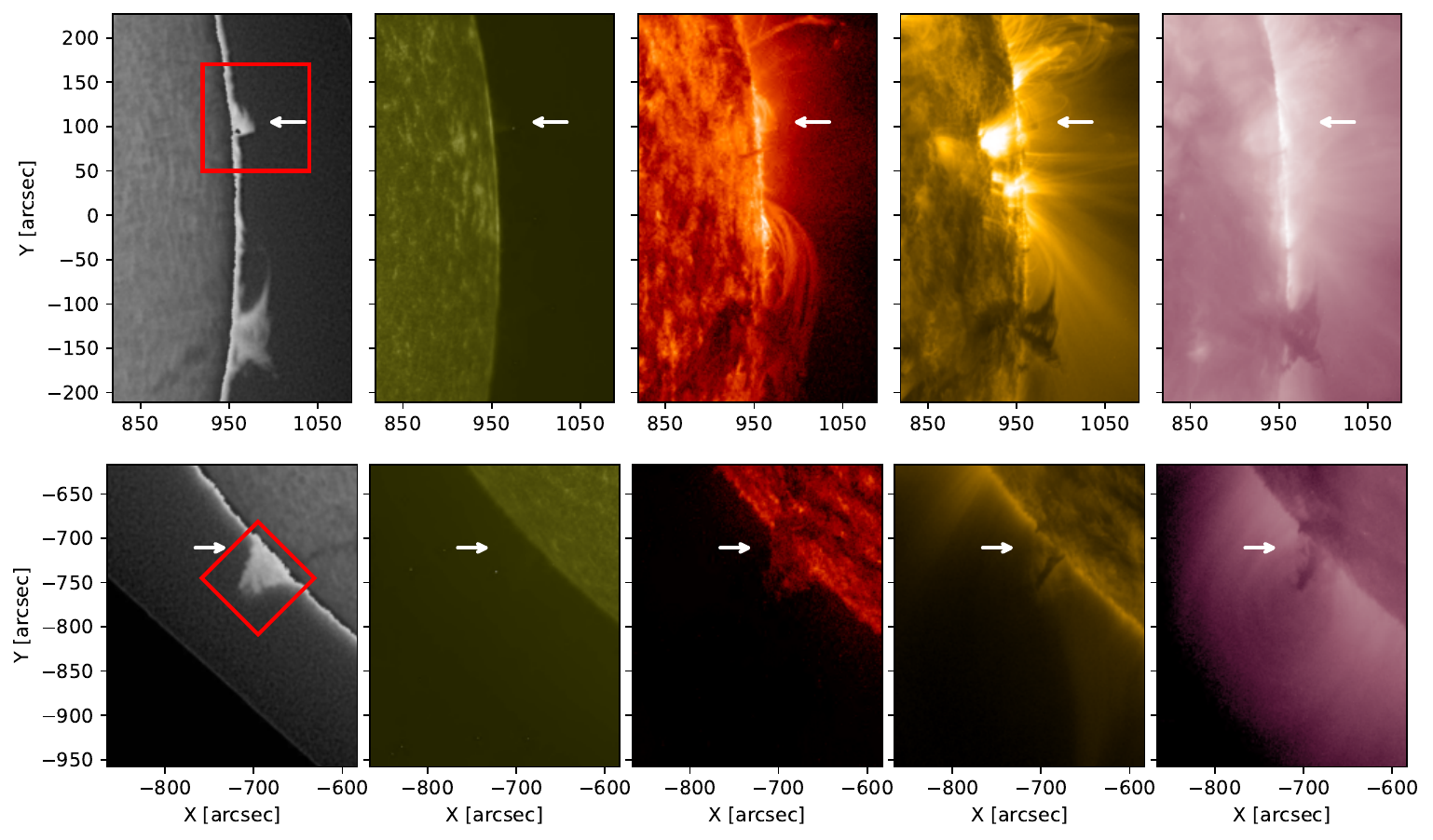}
\caption{GONG and SDO/AIA observations of two prominences in the H$\alpha$, 1600~\AA, 304~\AA, 171~\AA, 
    and 211~\AA\ channels, sampling plasma from the upper chromosphere and transition region to the low corona. The top row shows Prominence~1A (observed at 2025-10-09T08:47:53), located adjacent to an active region, while the bottom row shows Prominence~2 (observed at 2025-11-19T08:47:53) in a quieter coronal environment. White arrows indicate the prominence structures in each channel. The red boxes indicate the FOV of the cuts made in the figures below.}
\label{fig:promcontext}
\end{figure*}

Preliminary low-resolution spectra of prominences suggest that the signal originates primarily from emission by metals within the G-band region rather than from the CH lines. Therefore, referring to them as prominences in G-band is more appropriate than calling them G-band prominences. However, the low signal-to-noise of these observations does not rule out a contribution from CH absorption features. 

The present study aims to investigate this hypothesis in detail by reprocessing the original Sol'Ex data and making and analyzing additional observations with the Fast Multi-Line Universal Spectrograph (FaMuLUS, Denker et al. In prep.) camera system at the Echelle spectrograph of the 70-cm German Vacuum Tower Telescope \citep[VTT, ][]{Schroeter1985, vonderLuehe1998}. We also discuss the diagnostic potential of these lines.

\section{Observations and data processing}\label{observations}

\begin{figure*}[t]
%\centering
\includegraphics[width=1\textwidth]{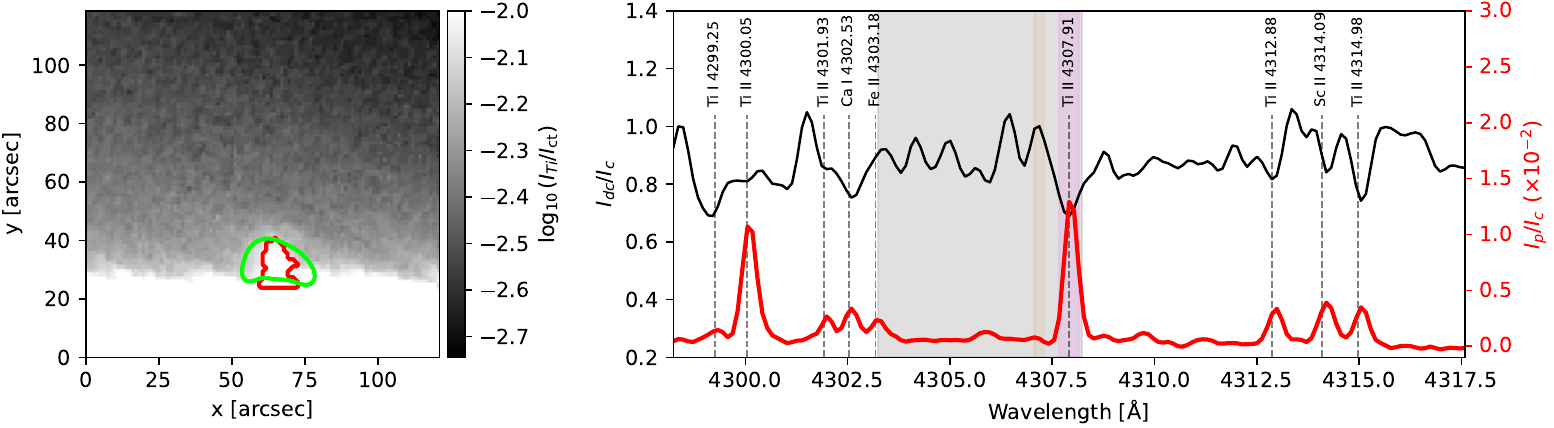}
\caption{Overview of Prominence~1A observations. \textbf{Left:} Prominence~1A imaged 
    in the \ion{Ti}{II}~4307.9~\AA\ line (violet line in right plot) shown in grayscale after the nearby continuum was subtracted (tan line in right plot). A green contour represents the extent of the prominence in a GONG H$\alpha$ filtergram.  \textbf{Right:} A disk center profile (black) showing the spectral extent of the recording, along with the prominence spectrum (red) after normalization to the disk center continuum intensity. The nine brightest emission lines are marked with dashed vertical lines. The gray shaded region corresponds to the spectral FaMuLUS window.}
\label{fig:solexprom}
\end{figure*}
\begin{figure*}
\includegraphics[width=1\linewidth]{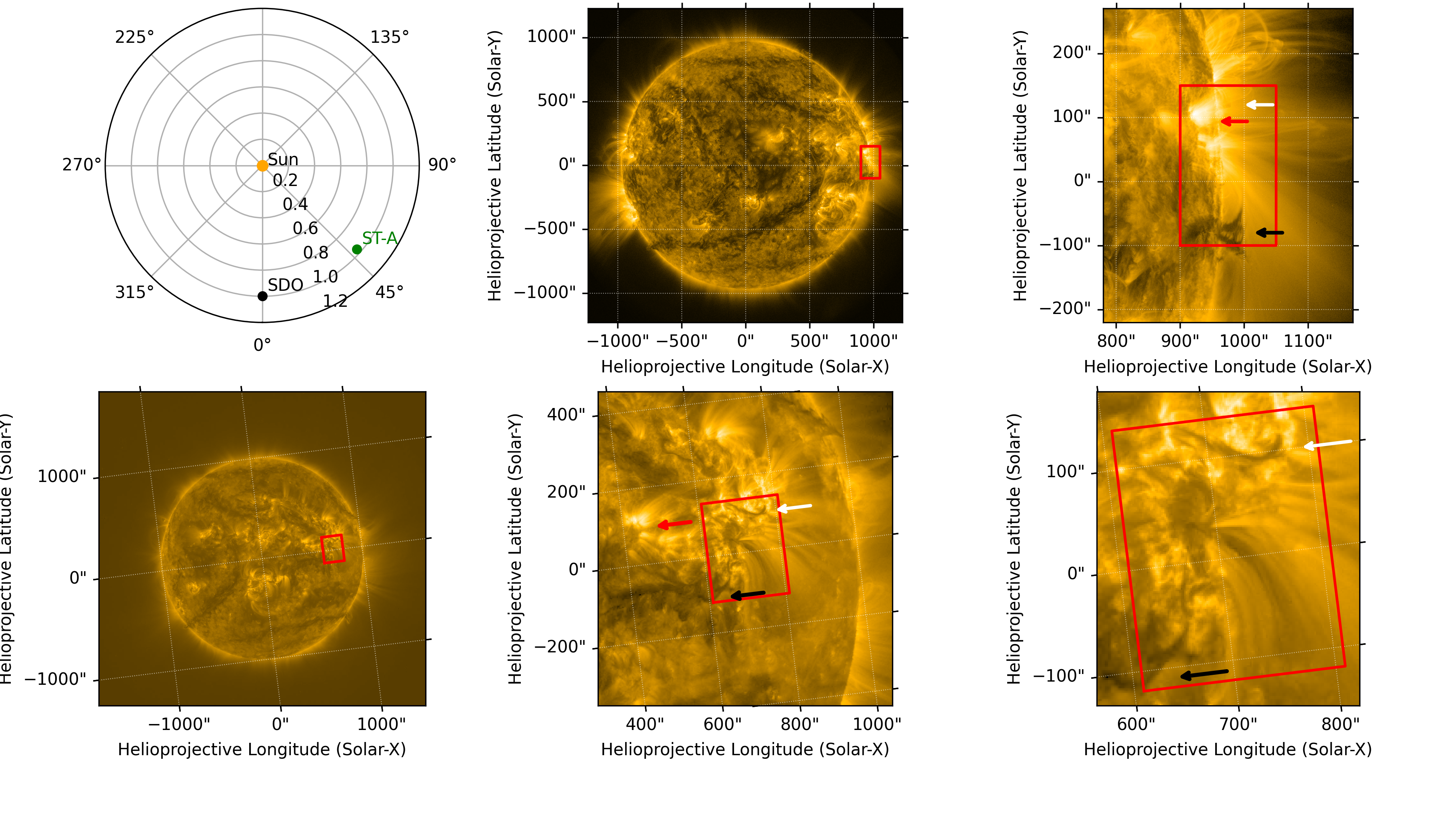}
\caption{
Viewing geometry and multi-scale context in the 171\,\AA\ channel.
\textbf{Top left:} Relative positions of SDO (near Earth) and STEREO-A in the heliographic Stonyhurst frame at the time of observation.
\textbf{Top middle:} Full-disk AIA 171\,\AA\ image with the selected limb segment shown in red.
\textbf{Top right:} Zoomed AIA view of the limb region containing the two prominences, with arrows marking the prominence footpoints.
\textbf{Bottom left:} Corresponding full-disk STEREO-A/EUVI 171\,\AA\ image with the projected limb segment.
\textbf{Bottom middle:} Intermediate EUVI zoom providing additional spatial context.
\textbf{Bottom right:} Tight EUVI zoom of the prominence region, with arrows indicating the same footpoints as in the AIA panel.
}
    \label{fig:EUI1}
\end{figure*}

Two prominences were observed in this work. For each one, we describe the observing instrument, the data processing workflow, and the prominence properties. Context images taken in H$\alpha$ with the Global Oscillation Network Group \citep[GONG,][]{Harvey1996} along with the 1600~\AA, 304~\AA, 171~\AA, and 211~\AA\ bands of the Atmospheric Imaging Assembly \citep[AIA,][]{Lemmen2012} on board of the Solar Dynamics Observatory \citep[SDO,][]{Pesnell2012} and shown in  Fig.~\ref{fig:promcontext}.

\subsection{Prominence~1A}

The observations of Prominence~1A were made with a modified Sol'Ex located in Cluj-Napoca, Romania. The spectroheliograph was attached to a 62/400 refractor, with an aperture stopped down to a diameter of about 4~cm, equipped with a 2-inch Altair solar contrast filter, which is centered at the G-band at 4303~\AA\ with a bandpass of 20~\AA\ and serves as an energy rejection filter. The Sol'Ex spectroheliograph has a grating constant of 2400 lines mm$^{-1}$, which corresponds to a spectral resolving power of ${\cal R} \approx 40\,000$ and can be tuned to a wide range of wavelengths \citep{Pal2025}. The instrument uses a monochromatic 16-bit Altair Hypercam 26M (APS-C) camera, which can capture weaker signals but at the same time reduces the spectral resolution by roughly a factor of two due to its large pixels. The setup can scan the full solar disk in around 4500 slit positions, taking about 1~min. For these observations, the Sol'Ex spectroheliograph was tuned to a spectral window extending from approximately 4298~\AA\ to 4317~\AA. 

The solar disk was reconstructed and stored as FITS data using the JSol’Ex\footnote{\href{https://github.com/melix/astro4j/tree/main/jsolex}{github.com/melix/astro4j/tree/main/jsolex}} open source software in the same way as described in \citep{Pal2025}. The wavelength and intensity calibration were performed with the HelioSpectrotron~5000 interactive atlas\footnote{\url{https://hs5000.vo.aip.de/}} \citep{Pietrow2026} in combination with the ISPy \citep{ISPy2021} spectral calibration library and the \citet{Neckel1984} solar spectral atlas.

Prominence~1A was observed on the west side of the limb at helioprojective Cartesian coordinates $(x,\, y) = (958\arcsec,\, 131\arcsec)$ from 08:29~UT to 11:00~UT on 2025 October~9 (See white arrow in Fig.~\ref{fig:promcontext}). During this time period, the prominence remained stable and displayed a relatively strong signal in the \ion{Ti}{II} line at 4307.9~\AA\ throughout the time series. We focus on the time period around 10:46~UT, because it had the best seeing.

Among all detections, Prominence~1A exhibits by far the strongest G-band signature observed with this instrument to date \citep{Varadi2025}. A clear contrast is provided by the prominence located south of Prominence~1A ($x,y = 960\arcsec, -150\arcsec$), as it appears larger and brighter in H$\alpha$ and looks more like a typical prominence in the EUV channels, it shows no detectable signature in the Sol'Ex data. We call this structure Prominence~1B.

Prominence~1A is detected in multiple AIA channels, but its emission is partly obscured by hot coronal plasma from active region NOAA~14232 that lies along the same line of sight. In AIA~304\,\AA, it appears comparatively brighter than Prominence~1B, indicating enhanced \ion{He}{II} emission. It is also visible as a narrow spine in AIA 1600~\AA, while no clear signature is detected in AIA 211~\AA. A dark triangular shape can be discerned in AIA 171~\AA\ behind the bright active-region loops. Nevertheless, the structure appears as an unambiguous prominence in the GONG H$\alpha$ observations.

The geometry of the system becomes clearer when viewed from the vantage point of the Solar Terrestrial Relation Observatory \citep[STEREO-A,][]{Howard2008} satellite, which is 48.5$^\circ$ degrees ahead of the Earth with respect to the heliographic frame (see top left panel in Fig.~\ref{fig:EUI1}). Here, the Extreme UltraViolet Imager \citep[EUVI,][]{Wuelser2004} observations demonstrate that the bright region is not physically connected to this structure (see red arrow in Fig.~\ref{fig:EUI1}). From this vantage point, Prominence~1A (white arrow) is found to be considerably darker and more compact than the nearby Prominence~1B (black arrow), and is surrounded by several small bright patches. Although this morphology makes a strict classification as either active or quiescent a matter of debate, its stability over a period of approximately 12~hours both before and after the observations suggests that, following the definitions of \citet{TandbergHanssen1995}, it is closer to a quiescent prominence.

However, the relatively strong signal of Prominence~1A in H$\alpha$, combined with its faint presence in the AIA~1700 channel, suggests that it is not independent of the surrounding active regions. Given that prominence emission is largely governed by resonant scattering of incident solar radiation \citep[e.g.,][]{Heinzel2015, Jenkins2023, Heinzel2025}, the bright surroundings of Prominence~1A mean it should not be considered quiet.

\subsection{Prominence~2}

The observations of Prominence~2 were made with the Fast Multi-Line Universal Spectrograph (FaMuLUS) camera system. This instrument is an upgrade of the Echelle spectrograph, around which the telescope was built \citep{Schmidt1991}, and consists of four independent CMOS cameras that can simultaneously observe between approximately 3900~\AA\ and 9000~\AA. The cameras cover spectral regions between approximately 4.6~\AA\ and 10.7~\AA, respectively. The respective spectral resolving power of the spectrograph extends from approximately ${\cal R} = 995\,000$ to 431\,000. The spectra are critically sampled at approximately 4950~\AA. At shorter wavelength the spectra are undersampled and at longer wavelength oversampled, with a  a pixel size of 4.6~$\mu$m and a 4-pixel binning along the dispersion direction.

For the present observations, FaMuLUS was configured to cover four spectral windows, that is, G-band at 4307~\AA, \ion{Cr}{I} at 5782~\AA, H$\alpha$ at 6563~\AA, and \ion{Ca}{II} at 8542~\AA. The respective exposure times were 300~ms, 400~ms, 970~ms, and 970~ms for each slit position. During one scan, which takes just over 6~min, the 216\arcsec-long slit scans a dense raster with 334 steps of 0.36\arcsec, resulting in slit reconstructed maps with a size of $216\arcsec \times 120\arcsec$. The data was reduced with the standard FaMuLUS data pipeline.

Prominence~2 was observed on the south-eastern side of the limb at helioprojective Cartesian coordinates $(x,\,y) = (-687\arcsec,\, -705\arcsec)$ at 08:25~UT on 2025 November~19 for one scan. Unlike Prominence~1A, this prominence is in a quiet-Sun region, far enough away from any active region, and can be considered a typical quiescent prominence \citep[e.g.,][]{Engvold2015}. Figure~\ref{fig:promcontext} shows a triangular shape in the AIA~304~\AA\ channel with a dense, narrow spine in the AIA 171~\AA\ channel. No signal is seen in the AIA 1600~\AA\ image, but an elliptical cavity, resembling the structure simulated and later modeled in \citet{Liakh2023} and \citet{Pietrow2024rot}, can be seen in the AIA 211~\AA\ channel. Its appearance in GONG resembles the more typical-looking Prominence~1B from the Sol'Ex observations.

Prominence~2 was also observed with Sol'Ex spectroheliograph, and a significantly weaker signal was found compared to Prominence~1A \citep{Varadi2025}, making it impossible to extract a usable spectrum from the 4-centimeter spectroheliograph. This means that Prominence 2 is likely more comparable to Prominence 1B than the active Prominence 1A.

\begin{figure*}[]
%\centering
\includegraphics[width=1\textwidth]{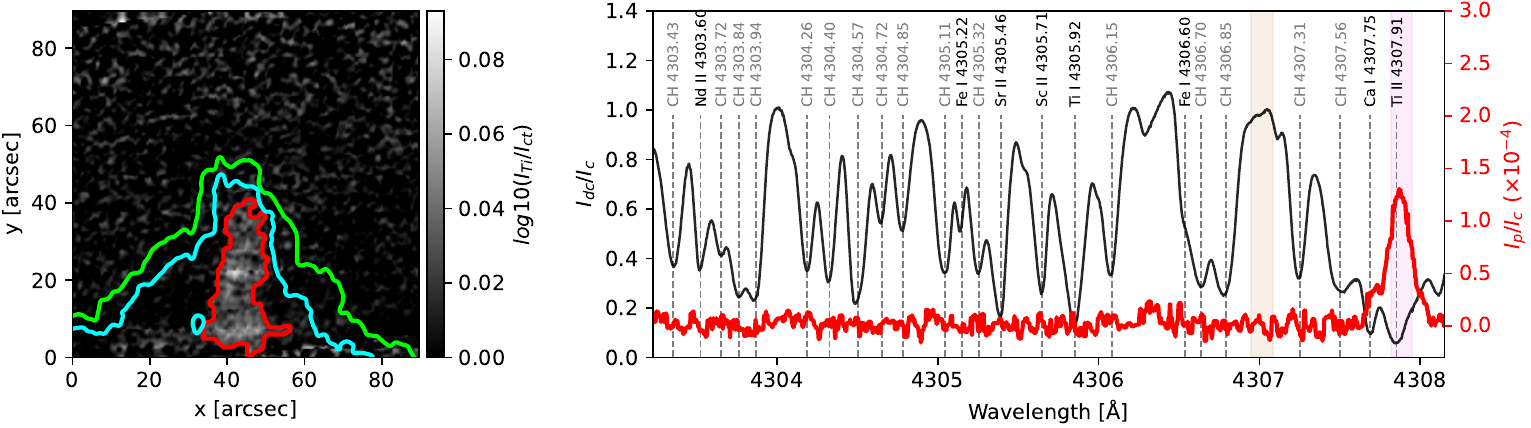}
\caption{Overview of Prominence~2 observations. \textbf{Left:} Prominence~2 imaged in the 
    \ion{Ti}{II} line at 4307.91~\AA (violet band in right figure), shown in grayscale as a logarithmic contrast plot with the nearby continuum (tan band in right figure). The green contour indicates the extent of the prominence in an H$\alpha$ slit-reconstructed line-core intensity map, the cyan contour traces the prominence extent in a \ion{Ca}{II} NIR slit-reconstructed line-core intensity map, and the red contour marks the prominence extent of the slit-reconstructed intensity map at 4307.91~\AA. \textbf{Right:} A disk-center profile (black) showing the spectral extent of the recording, along with the prominence spectrum (red) after background subtraction. Line identifications are denoted with  dashed vertical lines, black for atoms and gray for CH.}
\label{fig:famulusprom}
\vspace{0.1cm}
\end{figure*}

\section{Results and discussion}

In this section, we discuss the post-reduction methods for both prominences. Afterwards, we fit the FaMuLUS emission lines and estimate their Doppler broadening term. 

\subsection{Sol'Ex observations of Prominence~1A}

First, a second-order polynomial was fitted to the solar limb to quantify its curvature, and each image column was subsequently shifted to produce a flattened limb. The prominence signal was then isolated following \citet{Varadi2025} by averaging the three spectral pixels centered on the \ion{Ti}{II} 4307.9\,\AA\ line and subtracting the nearby continuum at 4307\,\AA, thereby constructing a filtergram that highlights the prominence structure. A contour was used to mark this structure in the spectral cube, and the same contour shifted by 20\arcsec\ to the left was used as a background reference. The final spectrum was obtained by subtracting these two structures and gauging their intensity relative to that of the quiet Sun. In the left panel of Fig.~\ref{fig:solexprom}, we show the resulting prominence signal, the fitted contour (red), and a comparative GONG H$\alpha$ contour (green). In the right panel of the same figure, we show the disk center spectrum in black, the resulting prominence spectrum in red, the \ion{Ti}{II} 4307.9~\AA\ average in violet, and the continuum in tan.

The resulting spectra show a relatively clean signal with nine emitting features, two of which are significantly stronger than the rest. Using \citet{Moore66} we identify the most likely candidates for these lines, although exact line identification remains difficult. The two strongest lines are attributed to  \ion{Ti}{II} at 4307.9~\AA\ and 4300.1~\AA. In addition, seven weaker emission features are attributed to \ion{Ca}{i}, \ion{Ti}{ii}, \ion{Fe}{ii}, and \ion{Sc}{ii} lines. Besides these nine lines, several additional features are visible between 4302~\AA\ and 4308~\AA\ in a spectral region with many CH features, but their signal and the spectral resolution of the instrument are both too low to attempt an identification. For this reason, the FaMuLUS window was selected to cover these wavelengths (denoted as a gray box in the right panel in Fig.~\ref{fig:solexprom}).

\subsection{FaMuLUS observations of Prominence~2}

A similar reduction strategy was used on the FaMuLUS data. First, an observation of a 1951 USAF resolution test chart (defined in MIL-STD-150A) was used to spatially align the spectral windows. Next, a second-order polynomial was fitted to a spectrally averaged G-band intensity map of the limb, and each column was shifted to create a straightened limb. 

Afterwards, an average was taken at the \ion{Ti}{II} line at 4307.91~\AA\ to isolate the prominence signal from the background (left panel in Fig.~\ref{fig:famulusprom}). The area within this prominence (red contour) is then applied to the G-band, H$\alpha$, and \ion{Ca}{II} data. The same mask was shifted to the left and right of the prominence and averaged to obtain a background estimate. The difference of these two averages can be found in the right panel of Fig.~\ref{fig:famulusprom} as a smoothed red profile. The same plot contains a black disk-center profile, where all lines with a significant depth are labeled based on the line list provided by \citet{Moore66}. 

This wavelength range overlaps with the shaded region in Fig.~\ref{fig:solexprom}. Within this interval, emission is detected only in the \ion{Ti}{II} line at 4307.91~\AA\ and in the \ion{Ca}{i} line at 4307.75~\AA, while no signal is observed in other spectral features. This implies that the G-band prominence signal is primarily associated with metal lines rather than lines of the CH molecule. 

\begin{figure*}
\centering
\includegraphics[width=1\textwidth]{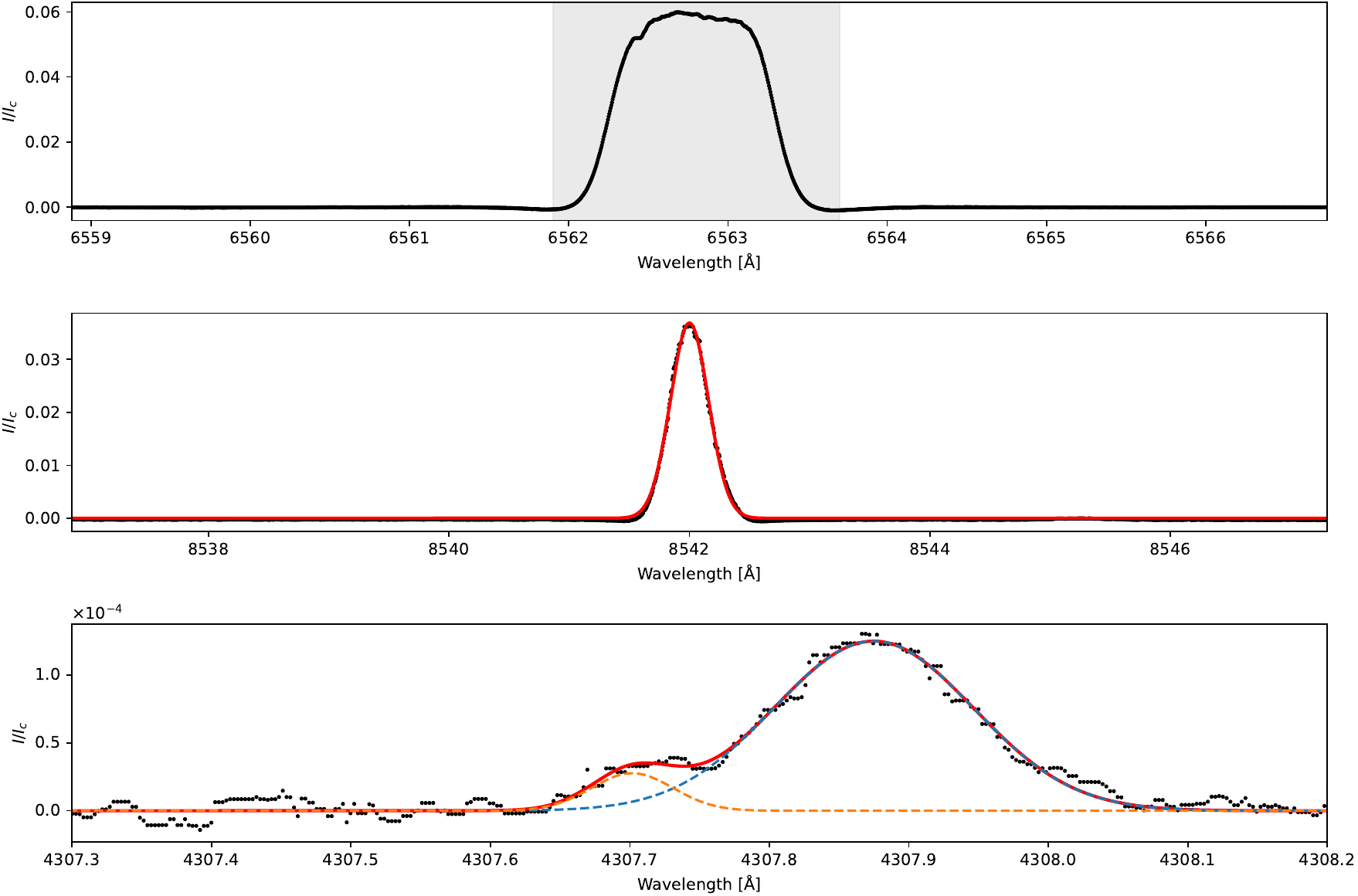}
\caption{Processed prominence signals in the H$\alpha$ line, the \ion{Ca}{II} line, and the 
    \ion{Ca}{I} and \ion{Ti}{II} line pair of the G-band, normalized to the local disk-center intensity. For the first row, a sum (gray box) is taken over the non-Gaussian emission profile to obtain the integrated line intensity of the prominence. A Gaussian was fitted to the emission of the other two prominences. The best fit parameters of these Gaussians are given in Table~\ref{tab:nb}.}
\label{fig:promint}
\end{figure*}

\begin{table*}
\centering
\caption{Summary of measured prominence line properties, including peak intensity, Gaussian width $\sigma$, line width $\Delta\lambda_{\mathrm{w}}$, and derived non-thermal velocities for assumed temperatures of 6000\,K and 8000\,K.}
\label{tab:line_props_wide}
%\centering
\begin{tabular}{lccccccc}
\hline\hline
Line
& $\lambda_0$
& $(I/I_c)_\mathrm{max}$
& $\sigma$
& $\Delta\lambda_w$
& $v^{2}$
& $v_{\mathrm{nth}}$ (8000~K)
& $v_{\mathrm{nth}}$ (6000~K)\rule[-5pt]{0pt}{16pt}\\
\hline

H$\alpha$
& 6562.79~\AA
& $6.00 \times 10^{-2}$
& -- 
& -- 
& -- 
& -- 
& --\rule{0pt}{11pt}\\

\ion{Ca}{II}
& 8542.09~\AA
& $3.68 \times 10^{-2}$
& 0.156~\AA
& 0.221~\AA
& 7.8~km$^{2}$~s$^{-2}$
& 2.1~km~s$^{-1}$
& 2.3~km~s$^{-1}$\rule[-5pt]{0pt}{11pt}\\

\ion{Ti}{II}
& 4307.91~\AA
& $1.25 \times 10^{-4}$
& 0.071~\AA
& 0.100~\AA
& 7.0~km$^{2}$~s$^{-2}$
& 2.0~km~s$^{-1}$
& 2.2~km~s$^{-1}$ \\

\ion{Ca}{I}
& 4307.75~\AA
& $2.76 \times 10^{-5}$
& 0.029~\AA
& 0.041~\AA
& 2.9~km$^2$~s$^{-2}$
& -- 
& 0.6~km~s$^{-1}$ \\

\hline
\end{tabular}
\label{tab:nb}
\end{table*}

\subsection{Diagnostic potential}

From the observations of Prominence~1A and Prominence~2, we find that the emission is dominated by \ion{Ca}{I} and \ion{Ti}{II} lines, with no discernible signal from the CH molecules. Using the NIST Atomic Spectra Database Lines Data \citep{NIST_ASD}, we find that both the \ion{Ca}{I} line at 4302.55~\AA\ and the \ion{Ca}{I} line at 4307.75~\AA\ are part of the same multiplet and share similar line strengths \citep{Olsen59}. For the \ion{Ti}{II} lines, we find two multiplets \citep{Pickering2001}, which connect the two bright lines and the four dimmer lines in Fig.~\ref{fig:solexprom}. 

The signal from Prominence~1A is approximately two orders of magnitude stronger than that of Prominence~2, which could be due to a combination of both physical and instrumental effects. Nevertheless, we restrict our quantitative analysis to Prominence~2 and use the Prominence~1A signal solely for line identification where possible.

Despite their relatively weak signals and blending, the combination of the \ion{Ca}{I} line at 4307.75~\AA\ and the \ion{Ti}{II} line at 4307.91~\AA\ makes this region an interesting diagnostic for comparisons between ions and neutrals. In particular, as a test of non-thermal motions between ions and neutrals, given that these lines are observed strictly simultaneously in the same wavelength range. 

A difference in these motions was first reported by \citet{Landman1981}, who demonstrated that spectral lines formed by ions and neutrals do not necessarily exhibit identical widths, as ions are directly coupled to the magnetic field, whereas neutrals are not. This suggests that neutrals will sink through the magnetic structure of the prominence when not in collisional equilibrium with charged particles, and such an effect was shown by \citet{Gilbert2002, Gilbert2007} for helium at the top of prominences. However, more recent work demonstrated that this is not always the case and neutrals have been reported with non-thermal widths similar to those of ions \citep[e.g.,][]{Stellmacher2015, Wiehr2019, Wiehr2021, Wiehr2025}. 

These works proposed an ionization memory effect, where recently recombined neutrals retain a memory of their prior ionized state, as the time between recombination and emission is much shorter than the collision time needed to erase the ion-induced velocity distribution. We repeated the methods of \citet{Wiehr2025} and applied them to the observations of Prominence~2. 

The intensities of the prominences are calibrated to the nearby disk-center continuum intensity using the ISPy library. In Fig.~\ref{fig:promint}, the signals of all three windows are shown in background-subtracted intensity units. The `total energy' of the prominence is calculated by integrating the H$\alpha$ line profile over wavelength (gray box in top panel of Fig.~\ref{fig:promint}), yielding the line-integrated specific intensity. This value is placed on an absolute scale using the quiet-Sun continuum from \citet{Labs2968}. The resulting line-integrated intensity is then compared to the values listed in \citet{Gouttebroze1993}, which can in turn be used to infer an approximate temperature for the prominence. However, this estimate is very sensitive to broadening effects induced by Doppler shifts, and given that we average over a large area (see Fig.~\ref{fig:famulusprom}), we likely integrate over several strands. This puts the temperature at either 6000~K or 8000~K. 

Following \citet{Wiehr2025}, we take the total broadening as the sum of the thermal and non-thermal terms $v^2_\mathrm{obs} = v^2_\mathrm{th} + v^2_\mathrm{nth}$. Here, the thermal term is defined as $v^2_\mathrm{th} = 2RT/\mu$, with $R$ the gas constant, $T$ the temperature, and $\mu$ the atomic mass. The total broadening is defined as $c\sqrt{2}\sigma/\lambda_0$, with $c$ the speed of light, $\sigma$ the width of the Gaussian, and $\lambda_0$ the rest wavelength. Combined this yields
\begin{equation}
C_{\mathrm{nth}}^{2} = \left( c\,\frac{\sqrt{2}\sigma}{\lambda_0} \right)^2
-
\frac{2 R T}{\mu}.
\end{equation}

For H$\alpha$, we have a maximum background-subtracted intensity of $6.00 \times 10^{-2} \rm I/I_c$ , and an integrated intensity of $5.99 \times 10^{-2} \AA\ \rm I / I_c$, which gives an energy of $1.68 \times 10^5$~erg~cm$^{-2}$~ster$^{-1}$~s$^{-1}$. In Table~\ref{tab:line_props_wide}, we show the line intensities normalized with respect to the local continuum, the fit parameters, and the non-thermal widths resulting from the two assumed temperatures. For 8000~K, we obtain a thermal width that is larger than the total width of \ion{Ca}{i}, which suggests that this temperature is too high for this prominence. At 6000~K, we find very similar non-thermal velocities for both ions but a much lower value for \ion{Ca}{i}, suggesting a more classical two-fluid approximation as suggested by \citet{Landman1981}. 

Given the low signal-to-noise ratio of this emission and the presence of significant blending from the nearby \ion{Ti}{II} line, no firm conclusions can be drawn for the \ion{Ca}{I} line at 4307.75~\AA\ line. Nevertheless, since most spectral lines examined in previous studies form higher in the atmosphere and correspond to more readily ionized species, it is plausible that the memory effect is less pronounced for weaker (low-forming in the photosphere) neutral lines such as the \ion{Ca}{I} line at 4307.75~\AA. If this is indeed the case, then such lines could be used together with more classically used ``prominence'' lines to accurately gauge the prominence temperature, similarly to how this is done with millimeter observations \citep{Labrosse2022}.

\section{Conclusions}

In this work, we presented spectroscopic observations of two solar prominences in the Fraunhofer G-band observed with an amateur-built Sol'Ex spectroheliopgraph and the FaMuLUS camera system at the VTT echelle spectrograph. In both cases, a clear emission was detected from ionized and neutral metal lines (primarily \ion{Ti}{II} and \ion{Ca}{I} lines) while no signal was measured in the plethora of molecular CH lines that dominate this wavelength window (see Figs.~\ref{fig:solexprom} and~\ref{fig:famulusprom}). In this sense, the observed structures are better described as prominences in the G-band, rather than ``G-band prominences'', as the latter would imply structures related to G-band diagnostics based on the CH molecule. 

We also demonstrate the value of collaborations with amateur astronomers who benefit from technological advances, making better telescope optics, filters, and spectrograph more affordable. In addition, amateurs have significant hands-on experience in solar observations, operate in large numbers, and in various locations, thus raising the odds of observing of rare events and exceptional features occurring on the Sun.

In these observations, we find no clear evidence for a strong ionization memory signature in the weak neutral line. If confirmed by follow-up observations with larger and more sensitive telescopes, this effect will make such lines a valuable diagnostic for constraining prominence temperatures.

\begin{acknowledgements}
AP was supported by grant PI~2102/1-1 from the Deutsche Forschungsgemeinschaft (DFG). MV acknowledges the support from IGSTC-WISER grant (IGSTC-05373). We thank Ralph Smith for his attempts to photometrically observe these prominences with an Altair G-Band Solar Filter. Additionally, we thank Aaron Peat and Veronika Jerčić for valuable discussions and comments on the manuscript. We also gratefully acknowledge the Struve Solar Society for fostering a collaborative space between amateur and professional solar observers.
The VTT is operated by a German consortium led by the Institute for Solar Physics (KIS) in Freiburg, with the AIP and the Max Planck Institute for Solar System Research (MPS) in G\"ottingen as partners. This research has made use of the bibliographic services of NASA's Astrophysics Data System (ADS). DeepL Write was used for assistance with grammar and language polishing.

\end{acknowledgements}

\bibliographystyle{aa}
\bibliography{ref}

\end{document}